\documentclass[RNAAS]{aastex62}
\usepackage{comment}
\usepackage{placeins}
\usepackage{xcolor}
\usepackage{csquotes}
\usepackage{href-ul}
\usepackage{float}

\begin{abstract}
    Gamma-ray bursts (GRBs) are the most energetic bursts of light in our universe, and rapid progenitor association of these events can lead to targeted and optimized follow-up observations, ultimately providing better insights about the physics involved. In this note, we investigate a semi-supervised machine learning algorithm, that utilizes label propagation, as a classification method. Using a dataset of 2512 GRBs we evaluate the method’s ability to assign probabilistic class memberships based on a subset of events with known progenitors. Further analysis is ongoing to improve the method and future progress will be made to refine the classification algorithm and the dataset.
\end{abstract}

\begin{document}

\title{Label Propagation for Identifying Gamma-Ray Burst Progenitors from Prompt Emission}

%\correspondingauthor{Nicol\'o Cibrario}
%\email{nicolo.cibrario@unito.it}
%\correspondingauthor{Michela Negro}
%\email{michelanegro@lsu.edu}
%\correspondingauthor{Skye Strain}
%\email{sstrai5@lsu.edu}
%\correspondingauthor{Eric Burns}
%\email{ericburns@lsu.edu}

\author[0009-0004-0546-0134]{Skye Strain}
\affiliation{Department of Physics \& Astronomy, Louisiana State University, Baton Rouge, LA 70803, USA}

\author[0000-0003-3842-4493]{Nicol\'o Cibrario}
\affiliation{Istituto Nazionale di Fisica Nucleare, Sezione di Torino, Via Pietro Giuria 1, 10125 Torino, Italy}%
\affiliation{Dipartimento di Fisica, Università degli Studi di Torino, Via Pietro Giuria 1, 10125 Torino, Italy}

\author[0000-0002-6548-5622]{Michela Negro}
\affiliation{Department of Physics \& Astronomy, Louisiana State University, Baton Rouge, LA 70803, USA}

\author[0000-0002-2942-3379]{Eric Burns}
\affiliation{Department of Physics \& Astronomy, Louisiana State University, Baton Rouge, LA 70803, USA}

%% Note that RNAAS manuscripts DO NOT have abstracts.
%% See the online documentation for the full list of available subject
%% keywords and the rules for their use.
% \keywords{editorials, notices --- 
% miscellaneous --- catalogs --- surveys}

%% Start the main body of the article. If no sections in the 
%% research note leave the \section call blank to make the title.
\section{Introduction} 

The classification of gamma-ray bursts (GRBs) by progenitor class remains a key goal in time-domain and multi-messenger astrophysics. Data-driven approaches, often based on machine learning, are increasingly used to investigate GRB origins from prompt emission. \cite{Negro2025} recently introduced a deep-learning framework employing a custom self-supervised pipeline and \textit{waterfall plots} as a novel data representation: 12 high-dimensional images encoding key prompt emission properties measured by the Fermi Gamma-Ray Burst Monitor (Fermi-GBM, \cite{GBM}). A set of autoencoders compresses these images into a compact 30-dimensional latent representation, which is then projected to two dimensions via UMAP \citep{UMAP} for visualization.

While \cite{Negro2025} demonstrated the framework's potential, two further steps are required to fully exploit it. First, the reliability of the 2D representation must be assessed, as addressed in \cite{2025RNAAS...9..302C} via a tailored implementation of the scDEED algorithm \citep{scdeed} that evaluates the 2D representation's robustness against the 30D one. The final step is assigning a classification label to each event, enabling estimation of the probability that a given burst belongs to one of the known GRB classes.

\section{The method}
 
Considering all detected GRBs with unknown progenitors, their classification would be an unsupervised machine learning problem, typically tackled with \textit{clustering} algorithms, that are able to identify sub-populations without prior labeling (e.g., \cite{Chattopadhyay_2017, Dimple_2023, Dimple_2024}).
However, since a fraction of detected GRBs has been confidently associated with known progenitors, we can instead formulate the problem within a \textit{semi-supervised} framework. Known GRBs serve as labeled examples, and we assign unlabeled events probabilities of belonging to each known class based on their distribution in the 30-dimensional representation. A widely used method is \textit{label propagation} \citep{label_propagation}, which exploits the data manifold structure and a limited set of labeled events to iteratively propagate class information, yielding probabilistic assignments for all events. We adopt the \href{https://sklearn.org/stable/modules/generated/sklearn.semi_supervised.LabelPropagation.html}{scikit-learn implementation}.
Label propagation is applied to the 30-dimensional latent representation from the autoencoder; the 2D UMAP embedding is used only for visualization and does not affect classification. Only GRBs with confirmed progenitors are included as labeled samples:

\begin{itemize}

    \item \textbf{GRBs from Massive Stars (MS)}: GRB~130215A, GRB~130427A, GRB~130702A, GRB~140606B, GRB~171010A, GRB~180728A, GRB~190114C, GRB~190829A, GRB~211023A, GRB~221009A, GRB~200826A, GRB~230812B\citep{cano2017observer, melandri2019grb, rossi2026grb, melandri2022supernova, bhirombhakdi2024redshift, belkin2021grb, blanchard2024jwst, ahumada2021discovery, hussenot2024multiband};
    
    \item \textbf{Neutron Star Mergers (NSM)}: GRB~150101B, GRB~160821B, GRB~170817A \citep{rastinejad2025uniform, abbott2017gravitational};
    
    \item \textbf{Magnetar Giant Flares (MGF)}: GRB~200415A, GRB~180128A, GRB~231115A \citep{svinkin2021bright, trigg2024grb, mereghetti2024magnetar};
    
    \item \textbf{Long GRBs from Mergers (LM)}: GRB~211211A, GRB~230307A \citep{rastinejad2022kilonova, levan2024heavy}.
    
\end{itemize}

The label propagation algorithm returns an $N \times k$ matrix, where $N$ is the total number of GRBs and $k=4$ is the number of known classes. Each row contains the posterior probabilities that a given event belongs to each class.

To incorporate prior knowledge on the expected class fractions, following \cite{label_propagation}, we re-weight the class masses inferred from label propagation. The class mass for class $k$ is defined as $m_k = \sum_i P_{ik}$, where $P_{ik}$ is the posterior probability of event $i$ belonging to class $k$. Given the expected class fraction $p_k$, we rescale each column as $P_{ik} \rightarrow P_{ik}' = P_{ik}\,(p_k / m_k)$. This enforces $\sum_i P_{ik}' \propto p_k$. Since this operation breaks row normalization, we renormalize each event as $P_{ik}^{\mathrm{final}} = P_{ik}' / \sum_k P_{ik}'$, preserving a proper probabilistic interpretation while imposing a soft prior on class fractions.

\section{Results}

We adopt theoretical detection rates as soft priors, specifically 60\% for MS, 19\% for NSM, 1\% for MGF, and 20\% for LM. These are based on the short–long GRB split from \cite{von_Kienlin_2020}, with the MGF fraction taken from \cite{Burns_2021}, and the NSM fraction defined as the remainder of the short GRB population. For long mergers, we adopt an median estimate from the range in literature of $\sim$5–50\%  of long GRBs from the literature \citep{Dado_2018, Xu_2025}. These are approximate, literature-informed values, that we use both as soft priors and as a reference baseline to assess the consistency of the inferred class fractions with current expectations.

We apply the method described in the previous section to our sample of 2512 GRBs and, for each event $i$, identify the progenitor class corresponding to the highest value of $P_{ik}^{\mathrm{final}}$. The resulting distribution shows that 78.34\% of events have the highest probability of originating from an MS progenitor, 18.83\% from NSM, 2.27\% from MGF, and 0.56\% from LM.

Comparing expected detection rates with our highest-probability fractions reveals discrepancies: MS events are overrepresented, while LM events are underrepresented. This likely reflects the structure of the 30-dimensional latent space, where these populations may partially overlap. Given the much larger number of known MS events, the algorithm may be biased toward assigning ambiguous events to the MS class, suppressing the LM population.
The probabilistic nature of the method, however, provides a more informative picture by quantifying classification confidence. By evaluating the fraction of total events with classification probability greater than 99\%, we find 20.66\% MS, 2.23\% NSM, 0.32\% MGF, and 0.40\% LM exceeding this threshold. These high-confidence subsets show that the algorithm can robustly identify a substantial portion of the population. The events not reaching such confidence, particularly within MS, are consistent with partially overlapping progenitor regions in the latent space. This highlights a key advantage of probabilistic classification: it identifies confident events while flagging ambiguous ones for cautious follow-up.

\begin{figure}[h]
    \centering
    \includegraphics[width=0.72\linewidth]{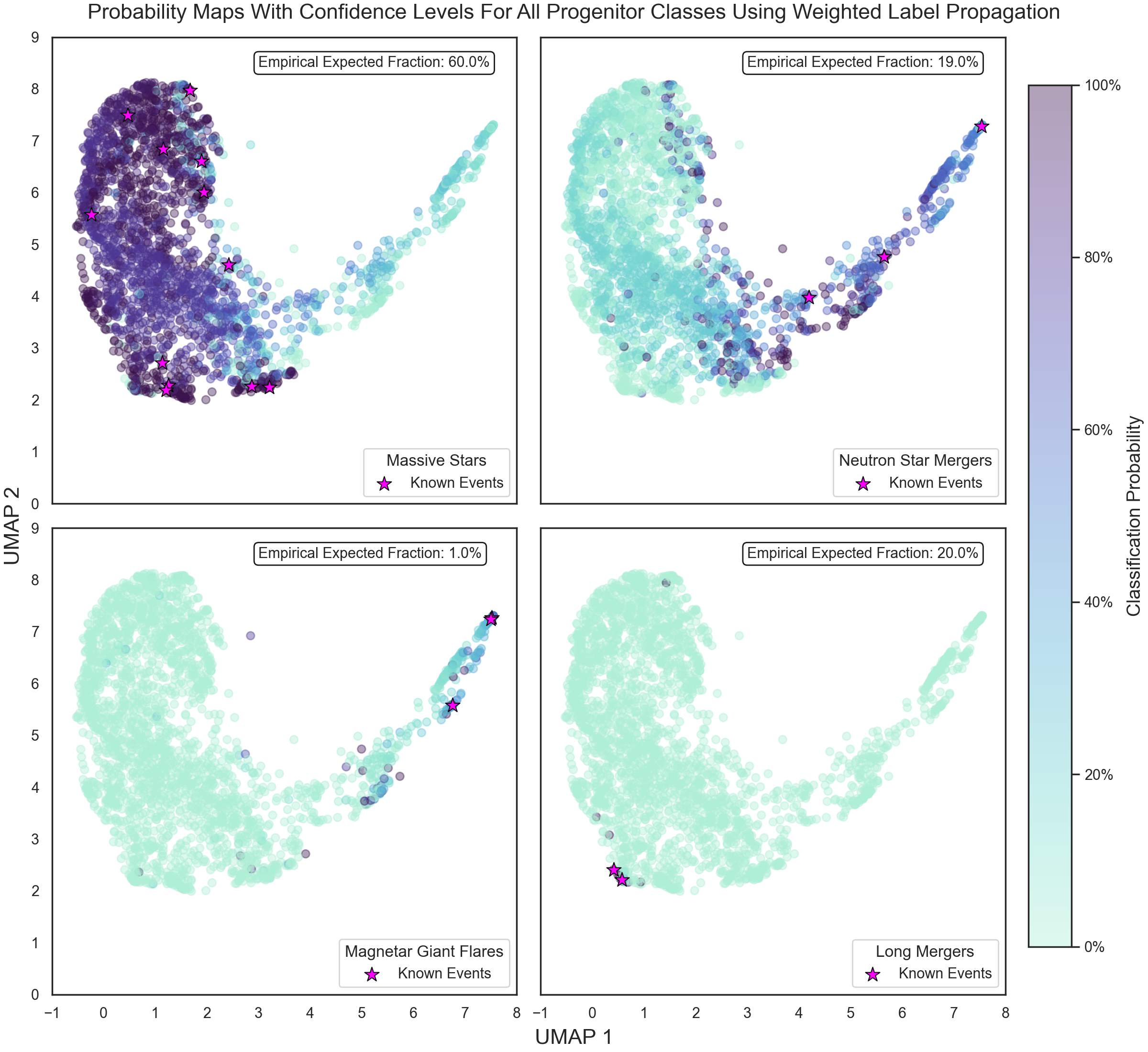}
    \caption{2D embedding of GRBs obtained with UMAP. Each panel corresponds to a different class (top left: MS, top right: NSM, bottom left: MGF, bottom right: LM), with the colorbar indicating the classification probability assigned by the label propagation algorithm after reweighting. Pink stars mark the known events used as the labeled sample during training. There is also a \href{https://grb-smartwaterfall-explorer.vercel.app/}{3D version} of this plot.}
    \label{fig:results}
\end{figure}

In Fig.~\ref{fig:results} we show the class probability distributions in the 2D UMAP embedding of the 30D space. We assessed how well this dimensionality reduction preserves local structure using the reliability technique of \cite{2025RNAAS...9..302C}. As discussed there, only a very small fraction of events ($<1\%$) have a non-reliable 2D representation, allowing us to treat the 2D embedding as a good lower-dimensional approximation of the 30D distribution. Since the embedding is used only for visualization, and marking non-reliable events would reduce figure readability, we refer the reader to \cite{2025RNAAS...9..302C} for details.

In conclusion, the current framework gives promising results while highlighting areas for improvement. Ongoing work focuses on improving input data quality, particularly through better background subtraction for waterfall plot generation, and on optimizing the machine learning architecture producing the 30D latent space. We also plan to add ultra-long GRBs as another known progenitor class, and to designate a subset of known GRBs as test events excluded from training, enabling performance evaluation against established classifications. Once implemented, our final goal is to integrate this study with previous works \citep{Negro2025, 2025RNAAS...9..302C} into a unified automated pipeline, enabling near real-time classification of newly detected Fermi-GBM GRBs with both a predicted progenitor class and an associated probability.

\bibliographystyle{aasjournal}
\bibliography{rnaas}
\end{document}